\documentstyle[12pt]{article}
\begin{document}
\vskip .27in
\begin{center}
{\large \bf Monte Carlo calculation of the spin-stiffness of the
two-dimensional
 Heisenberg model }\\
\ \\

Michel Caffarel\\
CNRS-Laboratoire de Dynamique des Interactions Mol\'eculaires,\\
 Universit\'e Paris VI,
 4 Place Jussieu 75252 Paris Cedex 05, France.\\
e-mail: mc@dim.jussieu.fr\\
\ \\

Patrick Azaria\\
Laboratoire de Physique Th\'eorique des Liquides\\
Universit\'e Paris VI,
 4 Place Jussieu 75230 Paris Cedex 05, France.\\
e-mail: aza@lptl.jussieu.fr\\
\ \\

Bertrand Delamotte and Dominique Mouhanna\\
Laboratoire de Physique Th\'eorique et Hautes Energies\\
 Universit\'e Paris VII, 2 Place Jussieu,
 75251 Paris Cedex 05, France.\\
e-mail: delamott@lpthe.jussieu.fr and mouhanna@lpthe.jussieu.fr\\
October 1993\\
\end{center}
\vskip 1cm

\begin{abstract}
Using a collective-mode Monte Carlo method (the Wolff-Swendsen-Wang algorithm),
 we compute the spin-stiffness
 of the two-dimensional classical Heisenberg model. We show that it is the
 relevant physical quantity
 to investigate the behaviour of the model in the very low temperature range
 inaccessible to  previous studies based on correlation length and
 susceptibility calculations.

PACS numbers: 11.10Gh,11.10Lm,75.10Hk,75.30Fv\\
\end{abstract}
\newpage

As well-known, the long-distance, low-energy physics of two-dimensional spin
 systems are
expected to be obtained from a low-temperature perturbative expansion of a
 suitable  Non Linear Sigma (NL$\sigma$) model.
However, the relevance of the low-temperature expansion beyond perturbation
 theory relies on the assumption of asymptotic freedom. Although this property
 is hardly questionable for symmetric $O(N)/O(N-1)$ models
with $N>2$ (see however \cite{1}) no definite answer is known for more general
 models, in particular for Non-Symmetric models such as $O(3)\otimes O(2)/O(2)$
 which are relevant for the study of frustrated Heisenberg spin systems
\cite{2}.
 Indeed, these models are suspected to be strongly affected by topological
 excitations \cite{3} which is
a possible source of failure of the low-temperature expansion. In order
to have a non perturbative control of the low-temperature expansion, one can
 take advantage of  Monte Carlo simulations. Up to now, calculations have been
 mainly concerned with  correlation lengths and susceptibilities \cite{4}.
 Unfortunately, because of their exponential behaviour as a function of
 $\beta=1/kT$ and the computationaly accessible lattice sizes, studying the
 very low temperature regime is very demanding, or even impossible.
The aim of this
 paper is to show that the relevant physical quantity allowing to reach this
 regime for accessible sizes is the spin-stiffness $\rho_s$, a measure of the
 free energy increment
under twisting of the boundary conditions \cite{5,6}.
 In the following we shall restrict ourselves to the two-dimensional classical
 Heisenberg model; applications to more involved models will be presented in a
 forthcoming work.

The Hamiltonian of the Heisenberg model is:
\begin{equation}
H=-J\sum_{<ij>}{\bf S}_i {\bf .} {\bf S}_j
\label{eq1}
\end{equation}
where $<ij>$ denotes
 the summation over nearest-neighbours of a finite square
 lattice of size $L$.
In (\ref{eq1}), ${\bf S}_i$ are three-component unit-length classical vectors
 and $J$ is positive. Each site $i$ of the lattice is indexed by two
 coordinates $x_i$ and $y_i$.

 We impose a twist
 in the ${\bf x}$ direction, by coupling the system with two walls of spins:
 ${\bf S(x=0)=S_1}$, ${\bf S(x=L)=S_2}$, ${\bf S_2}$  being deduced from
 ${\bf S_1}$  by a  rotation of angle $\theta$ around  a direction ${\bf e}$.
 The spin-stiffness $\rho_s$ is  defined as:

\begin{equation}
\rho_s(L)={\partial^2 F(\theta)\over \partial\theta^2} {\Biggr
\vert_{\theta=0}}\
 \ .
\label{eq2}
\end{equation}
where $F$ is the free energy.

In terms of the spins it writes:
\begin{equation}
\rho_s(L)= {J\over L^2}<\sum_{<ij>}({\bf S}_i.{\bf S}_j-{\bf S}_i.{\bf e}\ \
{\bf S}_j.{\bf e}
)\ \ (x_i-x_j)^2>-{J\over {T L^2}}<(\sum_{<ij>}({\bf S}_i\wedge
{\bf S}_j).{\bf e}\ \ (x_i-x_j)\ \ )^2> \ \
\label{eq3}
\end{equation}
where $T$ is the temperature and Boltzmann averages are performed with two
walls of parallel spins fixed at boundaries in the ${\bf x}$ direction.

The finite size behaviour of $\rho_s(L)$, when $L$ is much larger than the
 lattice spacing $a$ but much smaller than the correlation length $\xi$, has
 been calculated at one and two-loop order with use of the $O(3)/O(2)$
NL$\sigma$ model
 [5,7]:

\begin{equation}
\rho_s \sim {1\over 2\pi} \ln {\xi\over L} + {1\over 2\pi} \ln {\ln {\xi\over
L}}\ \
\label{eq4}
\end{equation}
\\
where the common coefficient $1/2\pi$ in front of the leading and sub-leading
 logarithmic terms
is a universal number which is not modified by higher orders in the
low-temperature expansion.

The crucial point in measuring $\rho_s$ is that its predicted size dependence
 given by (\ref{eq4}) is all
the more valid since $L\ll\xi$. Therefore, in the very low temperature regime
 we can hope to test formula (\ref{eq4}) by using a large range of
 relatively small lattice sizes. In contrast, measuring the
temperature dependence of $\xi$  requires  $\xi\leq L$ and therefore relatively
 high temperatures
 for accessible sizes\cite{4}, a regime where the validity of the perturbation
theory
 becomes less controlled.
 A most important point to notice is that at the very low temperatures
 considered here
 the physics of the model is entirely
 controlled by collective excitations - spin waves- and therefore we must
 take great care of these large-scale moves in any simulation of the
 model (``beating" the critical slowing down).

The purpose of this paper is to present a Monte Carlo study of the
 spin-stiffness for the
 finite two-dimensional classical Heisenberg model free of
 critical slowing down and then to investigate numerically
prediction (\ref{eq4}). To summarize what have been obtained, our Monte Carlo
calculations confirm the existence
of a leading logarithmic contribution with the universal amplitude $1/2\pi$.
In addition, an extra-contribution to the spin-stiffness consistent with the
subleading term of (\ref{eq4})
has also been clearly identified.
The Monte Carlo results presented have been obtained using the
 Wolff-Swendsen-Wang method \cite{8} of
updating large clusters of spins simultaneously. At the low
temperatures considered here, using a {\sl collective} Monte Carlo algorithm
 appeared to
be essential to get well-converged values of the spin-stiffness. In particular,
our preliminary attempts making use
of a Monte Carlo algorithm based on {\it local } spin updates
failed due to the severe critical slowing down.

To our knowledge, we present the first unambiguous numerical calculation
validating the precise
finite-size behavior of the spin-stiffness of the two-dimensional classical
Heisenberg model. It should be noted that a similar calculation has been
 reported recently
by Mon \cite{9}. However,
we disagree both
with the theoretical expression of the spin-stiffness used by the author
and with the relevance of the local Monte Carlo scheme employed in his work.\\

{\sl Results.}  The Wolff-Swendsen-Wang (WSW) algorithm has been implemented
 to simulate the
Heisenberg model on a $L \times L$ square lattice. In the y-direction periodic
boundary
conditions
have been chosen. In the x-direction, fixed boundary conditions are to be used.
Whereas in a local Monte Carlo algorithm imposing fixed walls of spins
pointing in some given direction is elementary, the situation is different
when clusters of spins are built with WSW. To escape
from this difficulty we have also chosen periodic boundary conditions
in the x-direction. This
introduces an error in the spin-stiffness exponentially small in $\ln L$.
As expected, this contribution
has been found to play no role in the following finite-size analysis.
We have found that relatively moderate sizes L are in fact sufficient to
validate formula (\ref{eq4}).
Lattices of sizes $L$=4,8,12,...32 have been simulated. We have performed our
 simulations at four
different temperatures: T/J= 0.1, 0.15, 0.3, and 0.395. In each case we are at
sufficiently
low temperature to be in the regime of validity of formula (\ref{eq4})
( L $ \ll \xi$).

Figure 1 presents the complete set of results obtained for the spin-stiffness
at different sizes and temperatures. At the scale of Figure 1, all curves
appear to be very rapidly linear as a function of $\ln L$. In order to
determine accurately the corresponding
slope a closer look is necessary. Figure 2 presents a blow up of data of
Figure 1 for the lowest (upper figure) and highest (lower figure) temperatures
treated,
$T/J=0.1$ and $T/J=0.395$, respectively. A first point to notice is that a very
high
accuracy on our data has been achieved. Such a level of accuracy is absolutely
necessary to
put into evidence the linear regime of the spin-stiffness as well as to get a
truly converged
estimate of the slope. We emphasize that only when resorting to a
collective Monte Carlo
scheme we have been able to fulfill both requirements. In our
first attempts to use
a local Monte Carlo scheme we observed
 a systematic and uncontrolled long-term drift of
the estimates of the statistical mean values.
A first important remark concerning Figure 2 is how fast we enter the linear
regime:
 at all temperatures considered  it is
reached at $L \sim 16$. By using data for $L=16,20,24,28,$ and 32 an estimate
of the slope
can be extracted, we get: -0.162(4),-0.166(5),-0.171(5), and -0.184(7)
at T/J=0.1, 0.15, 0.3, and 0.395, respectively.
At the very low temperature $T/J=0.1$ we recover within statistical
fluctuations
 the theoretical result $1/2\pi$ = 0.1592... predicted by formula (\ref{eq4}).
\cite{note1}
At higher temperatures non-negligible
higher-order contributions in the spin-stiffness show up. To put this on a more
quantitative basis, we
have performed a fit of the data using the full expression (\ref{eq4}).
The resulting curve is represented by a solid line in Figure 2.
The only free parameter entering the
fit is the correlation length $\xi$, the arbitrary reference value for the
spin-stiffness being
chosen so as to reproduce exactly the last data ($L=32$). The dashed line
is the linear curve obtained
when resorting to the leading logarithmic behaviour (no $\ln\ln$ corrections,
no renormalization
of the $1/2\pi$ slope) using the very same correlation length as determined in
the fit.
At $T/J=0.1$, both curves almost coincide in the linear regime, illustrating
the correctness
of the leading log prediction and the smallness of the higher-order corrections
at this temperature.
At the
higher temperatures considered, we clearly see the necessity of going beyond
leading order.
In addition, it is striking to see how good  representation (\ref{eq4}) is in
reproducing our
Monte Carlo data. Of course, at the accuracy determined by statistical
fluctuations it is
not realistic to hope to resolve the precise analytical $\ln\ln$ behavior of
the second-order
theoretical expression. However, our data are perfectly consistent with the
``renormalized slope"
predicted by (\ref{eq4}), $s^*= {\partial \rho_s}/{\partial \ln L} =
 -  {1}/{2\pi} (1 + 1/{\ln({\xi}/{L})})$.

In Figure 3 we have plotted the correlation length $\xi$ issued from the fit
using formula (\ref{eq4}). We also present the curve obtained from the
formula proposed by S.H. Shenker and J. Tobochnik \cite{10}
(obtained by matching high- and low-temperature calculations):
\begin{equation}
\xi \simeq 0.01 \;\; \frac{ \exp {2\pi J/T}}{1 + 2\pi J/T}
\label{eq5}
\end{equation}
It is very satisfactory to see that our rough estimates of $\xi$ are
in good agreement with this
completely independent
calculation of the correlation length.\\

\newpage
{\large \bf Figure Captions}\\

{\bf Fig}.1  Spin-stiffness for different sizes and temperatures. Statistical
fluctuations smaller than the size of crosses.\\

{\bf Fig}.2 Blow up of Fig.1 for $T/J=0.1$ and $T/J=0.395.$ The solid
 line is the
best fit using Eq.\ref{eq4}, the dashed line the first-order prediction (no
renormalization of the slope).\\

{\bf Fig}.3 Correlation length $\xi$. The solid line is obtained from
 Eq.\ref{eq5}, the values indicated by crosses from the fit of our
 data using Eq.\ref{eq4}.\\

\begin{thebibliography}{99}
\bibitem{note1} In fact, at this temperature the slope is very slightly
renormalized. Using Eqs.\ref{eq4}-\ref{eq5} we get -0.162 instead of
the bare value of -0.1592... However, both values are
 not distinguishable within statistical fluctuations.
\end{thebibliography}
\end{document}